# Characterization of Zero-point Vibration in One-Component Crystals

Yong Yang[1, 2]* and Yoshiyuki Kawazoe[2]

1. Water South Street, Cheng Xiang, Xing Bin Qu, Lai Bin, Guang Xi 546101, China
2. Institute for Materials Research, Tohoku University, Sendai 980-8577, Japan

We study the magnitude of zero-point vibration in one-component crystals. For the crystals whose constituent atoms share the same bonding geometry, we prove the existence of a characteristic temperature, $T_0$, at which the magnitude of zero-point vibrations equals to that of the excited vibrations. Within the Debye model $T_0$ is found to be ~1/3 of the Debye temperature. The results are demonstrated in realistic systems.

PACS numbers: 63.20.-e, 63.20.dk, 63.10.+a

**\*** Present address: National Institute for Materials Science, Tsukuba, Japan.

E-mail: yyangtaoism@gmail.com

***Introduction*** – Zero-point vibration (and the related zero-point energy) is one of the most remarkable effects predicted by quantum theory. A famous example is the ground state of the quantum harmonic oscillator, one of the few exactly solvable theoretical models for the Schrödinger equation. On the experimental side, one fact supporting the existence of zero-point vibration is the non-freezing behavior of liquid helium at ambient pressure when the temperature approaches the absolute zero (0 K). Another phenomenon serving as the evidence of zero-point energy is the Casimir effect [1], which predicts the existence of a weak force between two parallel metallic plates, and is confirmed by experiments in recent decades [2-4], though there is still some debate on the explanation of the origin of the force as zero-point energy in vacuum [5]. Generally, it is a common sense that quantum phenomena such as zero-point vibration are significant mainly in physical systems operating at cryogenic temperatures near the absolute zero.

More recently, however, one optical experiment has reported room-temperature quantum entanglement in the atomic vibrations of two macroscopically separated diamond crystals, in which the atomic vibration at ground state, i.e., the zero-point vibration plays a key role [6]. This experiment indicates a way for detecting room-temperature quantum phenomena at diamond-based systems. Then, the problem is, in addition to diamond, what other systems may be the potential candidates for the playground of measuring macroscopic quantum phenomena at temperatures well above 0 K? Or, for a given system, how to measure the role of zero-point vibration in the total atomic vibrations at the temperature of interest?

In this work, we attempt to provide a generalized approach for characterizing the magnitude of zero-point vibration in one-component crystals (consisting of one type of atoms) as a function of temperature. Given that the bonding geometry of each atom is identical, we can show the existence of a characteristic temperature, $T_0$, at which the atomic zero-point vibration and the excited vibrations are of the same magnitude. Below $T_0$, zero-point vibration plays a dominant role. Within the Debye model, we are able to establish a simple relation that $T_0 \sim 1/3\, \theta_D$, with $\theta_D$ being the Debye temperature. Calculations based on first-principles show that, for the materials with a high Debye temperature, the zero-point vibration is important not only at low temperatures, but also at the room temperature. The physical conclusions of this paper derive from some mathematical results will be stated in the form of two theorems. The proofs for the theorems will be given in the appendix.

*General Formalism* – We consider a one-component crystal consisting of $N$ primitive cells. For the $j$th atom in the primitive cell, within the harmonic approximation, the displacement from its equilibrium position can be expressed as follows [7]:

$$\vec{u}(j) = \sum_{q,s} (\frac{\hbar}{2Nm\omega_{q,s}})^{\frac{1}{2}} \vec{e}_{q,s}(j)(a_{q,s} e^{i\vec{q}\cdot\vec{R}} + a^+_{q,s} e^{-i\vec{q}\cdot\vec{R}}) \equiv \sum_{q,s} \vec{u}_{q,s}(j), \quad (1)$$

where the number of atoms per primitive cell is $r$, and $j = 1, \ldots, r$. $\hbar$ is the reduced Planck constant; $m$ is the mass of each atom; $\vec{q}$ is the wave vector in reciprocal space and $\vec{R}$ is the translational lattice vector denoting the position of the atom. The vibrational mode of the $s$th ($s$ = 1, 2, .., $3r$) phonon branch with a wave number $q\,(=|\vec{q}|)$ and a frequency $\omega_{q,s}$ is denoted by $(q, s)$ hereafter; $\vec{e}_{q,s}(j)$ is the phonon polarization vector. $a^+_{q,s}$ and $a_{q,s}$ are respectively the creation and annihilation operator for phonons. Obviously, $\vec{u}_{q,s}(j)$ represents the contribution from each mode $(q, s)$. Instead of $\vec{u}(j)$, the quantity can be measured experimentally is the mean square displacement (MSD), $|\vec{u}(j)|^2$. At the eigenstate $|n\rangle$ of vibrational mode $(q, s)$, the MSD is

$$\langle n|\vec{u}_{q,s}(j)\cdot\vec{u}_{q,s}(j)|n\rangle \equiv \langle u_n^2(j)\rangle_{q,s} = (n+\frac{1}{2})(\frac{\hbar}{Nm\omega_{q,s}})|\vec{e}_{q,s}(j)|^2.$$

The total MSD $\langle u^2(j) \rangle$ for each atom is contributed from all the vibrational modes and can be measured by the Debye-Waller factor (DWF) [8, 9] defined in X-ray diffraction [10]. In a canonical ensemble, the probability of finding the system at the eigenstate $|n\rangle$ of mode $(q, s)$ is

$$w_n(q,s) = \exp[-\beta E_n(q,s)]/Z, \text{ where } E_n(q,s) = (n+\frac{1}{2})\hbar\omega_{q,s}, \quad Z = \sum_n \exp[-\beta E_n(q,s)], \text{ and}$$

$\beta = 1/(k_B T)$. $k_B$ is the Boltzmann constant and $T$ is temperature. The weight factor is calculated to be $w_n(q,s) = (1 - e^{-\beta\hbar\omega_{q,s}})e^{-\beta n\hbar\omega_{q,s}}$. The MSD contributed from mode $(q, s)$ is then a weighted sum of all the eigenstates:

$$\langle u^2(j) \rangle_{q,s} = \sum_n w_n(q,s) \langle u_n^2(j) \rangle_{q,s} = \sum_n w_n(q,s)(n+\frac{1}{2})(\frac{\hbar}{Nm\omega_{q,s}})|\vec{e}_{q,s}(j)|^2, \quad (2)$$

for the $j$th atom in the primitive cell. Consequently, the total MSD is given by summing over all the vibrational modes:

$$\langle u^2(j) \rangle = \sum_{q,s} \langle u^2(j) \rangle_{q,s}. \quad (3)$$

The total MSD can be decomposed into two terms: $\langle u^2(j) \rangle = \langle u^2(j) \rangle_0 + \langle u^2(j) \rangle_T$, where $\langle u^2(j) \rangle_0$ is solely from the zero-point vibrational state $|0\rangle$ and $\langle u^2(j) \rangle_T$ from all the excited vibrational states $|n\rangle$ ($n \geq 1$). That is, $\langle u^2(j) \rangle_0 = \sum_{q,s} w_0(q,s) \langle u_0^2(j) \rangle_{q,s}$, and $\langle u^2(j) \rangle_T = \sum_{q,s} \sum_{n\geq 1} w_n(q,s) \langle u_n^2(j) \rangle_{q,s}$. For a mode $(q, s)$, the MSD in Eq. (2) can also be written as the sum of another two terms: $\langle u^2(j) \rangle_{q,s} = \frac{\hbar}{Nm\omega_{q,s}}(\sum_n n \cdot w_n(q,s) + \sum_n \frac{1}{2} \cdot w_n(q,s))$. Using $\sum_n w_n(q,s) = 1$, the term is reduced to: $\langle u^2(j) \rangle_{q,s} = \frac{\hbar}{Nm\omega_{q,s}}(\frac{1}{e^{\beta\hbar\omega_{q,s}}-1}+\frac{1}{2}) \equiv \frac{\hbar}{Nm\omega_{q,s}}(\bar{n}+\frac{1}{2})$,

where $\bar{n}$ is just the average number of phonons at temperature $T$ from the Bose-Einstein statistics.

For a macroscopic crystal in which the number of primitive cells $N$ is large enough, the values of wave number $q$ can be viewed as continuous, and summation in the $q$-space can be replaced by an integral:

$$\langle u^2(j) \rangle = \sum_{q,s} \langle u^2(j) \rangle_{q,s} = \int_0^{\omega_m} |\vec{e}_\omega(j)|^2 \frac{\hbar}{Nm\omega}(\frac{1}{e^{\beta\hbar\omega}-1}+\frac{1}{2})G(\omega)d\omega, \qquad (4)$$

where the term $|\vec{e}_{q,s}(j)|^2$ in Eq. (2) is replace by $|\vec{e}_\omega(j)|^2$, because $|\vec{e}_{q,s}(j)|^2$ is the same for the same frequency $\omega$. The vibrational density of states (VDOS) $G(\omega)$ satisfies $\int_0^{\omega_m} G(\omega)d\omega = 3Nr$, with $r$ the number of atoms per primitive cell, and $\omega_m$ the maximum of frequency.

For a one-component crystal, if the bonding geometry of each constituent atom is identical (referred as *geometric condition* hereafter), it follows that the term $|\vec{e}_{q,s}(j)|^2 = \frac{1}{r}$ (Proved in the Lemma of Appendix A.), and despite the difference by a phase factor, the magnitude of vibrations for each atom is the same. Consequently, the notations for the total MSD, and the MSD from zero-point vibration and the excited vibrations can be simplified as $\langle u^2 \rangle$, $\langle u^2 \rangle_0$, and $\langle u^2 \rangle_T$, respectively. Moreover, the vibrations of each atom are described by the same VDOS, which is $g(\omega) = G(\omega)/(Nr)$, where $r = 1$ for crystals with monoatomic basis and $r \geq 2$ for crystals with multiatomic basis. The corresponding MSD quantities are as follows:

$$\langle u^2 \rangle = \int_0^{\omega_m} \frac{\hbar}{m\omega} (\frac{1}{e^{\beta\hbar\omega}-1} + \frac{1}{2}) g(\omega) d\omega, \quad (5)$$

$$\langle u^2 \rangle_0 = \int_0^{\omega_m} \frac{\hbar}{2m\omega} (1 - e^{-\beta\hbar\omega}) g(\omega) d\omega, \quad (6)$$

$$\langle u^2 \rangle_T = \int_0^{\omega_m} \sum_{n \geq 1} (1 - e^{-\beta\hbar\omega}) e^{-\beta n\hbar\omega} (n + \frac{1}{2}) \frac{\hbar}{m\omega} g(\omega) d\omega, \quad (7)$$

where $\int_0^{\omega_m} g(\omega) d\omega = 3$.

From the expressions (see also in Appendix A.), it is found that $\langle u^2 \rangle_0$ decreases monotonically with temperature $T$ while $\langle u^2 \rangle$ and $\langle u^2 \rangle_T$ show opposite behavior. Moreover, under the geometric condition, we can state the following:

*Theorem 1.* – (Proved in Appendix A.) There exists a characteristic temperature, $T_0$, at which the magnitude of zero-point vibration is equal to that of the sum of excited vibrations, i.e., $\langle u^2 \rangle_0 = \langle u^2 \rangle_T$. Below $T_0$, $\langle u^2 \rangle_0 > \langle u^2 \rangle_T$.

*Theorem 2.* – (Proved in Appendix B.) Within the Debye model, there is simple relation that $T_0 = \theta_D / \zeta$, where $\theta_D$ is the Debye temperature, and $\zeta \approx 2.99345...$

*Realistic Applications* – We go on to demonstrate the two theorems mentioned above in two realistic systems that satisfy the geometric condition: aluminum (Al) and diamond, whose primitive cell contains one and two atoms, respectively. The VDOS for each atom, $g(\omega)$, is computed by using the program Quantum ESPRESSO [11], which is based on density functional perturbation theory (DFPT) [12]. The wave function of valence electrons is expanded using a plane wave basis set with a kinetic energy cut-off of 20 Ry for Al and 27 Ry for diamond. The ion-electron interactions are described by the von Barth-Car pseudopotentials [13], and the exchange-correlation energies are described using the Perdew-Zunger functional [14]. For the calculation of electronic wave function, the Brillouin zone is sampled using a $20 \times 20 \times 20$ Monkhorst-Pack k-mesh [15]. A $16 \times 16 \times 16$ uniform $\vec{q}$-point grid is used for the calculation of dynamical matrices, which are then diagonalized to get the eigenfrequency $\omega(\vec{q})$ ($\vec{q}$: wave vector). The dynamical matrices of denser $\vec{q}$-grids can be obtained by using interpolation method.

Figure 1 shows the calculated VDOS of Al (Fig. 1(a)) and diamond (Fig. 1(b)), whose features compare well with experimental measurements (Al) [16] or previous theoretical works (diamond) [17]. The maximum frequency is $\omega_m \approx 2\pi \times 1.016 \times 10^{13}$ rad/s (wave number: $\tilde{\nu}_m = 338.9$ cm$^{-1}$) for Al, and $\omega_m \approx 2\pi \times 3.990 \times 10^{13}$ rad/s ($\tilde{\nu}_m = 1331$ cm$^{-1}$) for diamond. The corresponding VDOS from the Debye model ($g(\omega) \propto \omega^2$) for Al and diamond are shown along with the DFPT ones, where the Debye frequency $\omega_D = k_B \theta_D / \hbar$, and $\theta_D = 428$ K for Al [18] and $\theta_D = 2230$ K for diamond [18]. It is clear that the Debye VDOS matches well with that of the DFPT method at the low frequency part.

The MSD quantities are shown in Figs. 1(c)-(d), for calculations using the VDOS from DFPT method and the Debye model. As expected above, $\langle u^2 \rangle_0$ decreases monotonically with increasing temperature while the terms $\langle u^2 \rangle$ and $\langle u^2 \rangle_T$ show opposite behavior. Using the DFPT VDOS, the characteristic temperature $T_0$ is determined to be ~ 148 K and ~ 687 K for Al and diamond, respectively. Within the Debye model, the value of $T_0$ turns out to be ~ 143 K for Al and ~ 745 K for diamond. It can be easily checked that the relation $T_0 \approx \theta_D / 2.99345$ stated in theorem 2 is well satisfied for the $T_0$ obtained using the Debye VDOS. The calculated total MSD at $T = 4.2$ K, 77 K, and 293 K, are listed in Table I for both systems, with comparison to that from the literature [10]. One finds satisfactory agreement, despite the systematic deviation between the DFPT and Debye data lines, which is due to the different VDOS employed in calculation.

The values of $T_0$ imply that, the magnitude of zero-point vibration of both Al and diamond is not negligible even at room temperature. Indeed, at $T = 298$ K, the ratio $\eta(T) = \langle u^2 \rangle_0 / \langle u^2 \rangle$ from DFPT is ~ 20.6% for Al and ~ 86.5% for diamond. Considering the fact that thermal expansion is caused by the anharmonic motions of atoms [18], and the small thermal expansion coefficients of both Al (~ 23 × $10^{-6}$/K) and diamond (~ 1 × $10^{-6}$/K) [17, 19], the anharmonic contribution to the total MSD is negligible at room temperature and below. Therefore, the results regarding the contribution of zero-point vibration to total MSD should preserve, and for diamond zero-point vibration plays a dominant role at room temperature. This is consistent with the recent experimental work on diamond-based system [6].

We have also performed calculations on a number of other elements whose crystals satisfy the geometric condition mentioned above. The corresponding values of $T_0$ for Be, Ti, Ta, and Pb are given in Table II, along with that of Al. It is clear that relation $T_0 \approx \theta_D / 2.99345$ holds precisely. From Fig. 1, in spite of the notable difference in the VDOS at the high frequency part, the values of $T_0$ don`t differ significantly between DFPT and Debye model. Therefore, the exact result $T_0 \approx \theta_D / 2.99345$ derived within the Debye model, is a good approximation for characterizing the magnitude of zero-point vibration in one-component crystals that satisfy the geometric condition, which is true for the crystals of most (if not all) elements in the periodic table. As a result, one can expect that, besides diamond, for the crystals of other elements with a high Debye temperature, such as beryllium ($\theta_D$ = 1440 K) [18], $T_0$ will also be a high value. Zero-point vibration in such systems will play a dominant role even at the room temperature. On the other hand, for most conventional superconducting elements whose superconductivity is mediated by electron-phonon interaction, e.g., the ones listed in Table II, the values of $T_0$ are well above the superconducting transition temperature $T_c$. As seen in Table II, $\eta(T_c) = \langle u^2 \rangle_0 / \langle u^2 \rangle \approx 1$ for all the elements enumerated there, which demonstrates that zero-point vibration of atoms plays a major role in conventional superconductivity. Indeed, in the time-tested BCS theory, only zero-point vibration is considered in the calculation of matrix elements for the electron-phonon interaction [20]. Though this scheme is intuitively reasonable, our work provides a solid theoretical justification for such approximation.

*Concluding Remarks* – In summary, the magnitude of atomic zero-point vibration as well as excited vibrations in one-component crystals is characterized theoretically. For crystals whose constituent atoms have the same bonding geometry, there exists a characteristic temperature $T_0$, at which the mean square displacement (MSD) of zero-point and excited vibrations has equal magnitude. Below $T_0$ the zero-point vibration is dominant over the excited vibrations. Within the Debye model, we obtain a simple relation between $T_0$ and the Debye temperature. The results can be briefly summarized in two theorems, which are further demonstrated in realistic systems by numerical simulations. In the case of Al and diamond, it is found that zero-point vibration plays a nontrivial role at relatively elevated temperatures. The role of zero-point vibration in the conventional superconductivity is discussed. Possible directions for the future research include extension of the present work to low-dimensional systems, such as nanotube (one-dimensional) and graphene (two-dimensional), and to the more complex systems whose constituent atoms have different bonding geometries.

**Acknowledgements**: The authors would like to express their sincere thanks to the crew of Center for Computational Materials Science of the Institute for Materials Research, Tohoku University, for their continuous support of the SR11000 supercomputing facilities.

## APPENDIX A

To prove theorem 1, we first establish the following:

*Lemma.* – For a one-component crystal, if all the atoms share the same bonding geometry (*geometric condition*), then the square of modulus of each polarization vector is identical, that is:

$$|\vec{e}_{q,s}(j)|^2 = \frac{1}{r}, \qquad (A.1)$$

where $r$ is number of atoms per primitive cell, and $j = 1, \ldots, r$.

*Proof.* – Atomic vibrations of a crystal can be studied within the primitive cell, by using the translational geometry. The vibrational properties are completely determined by the dynamical matrix $D(\vec{q}) = (D_{\alpha\beta}(\genfrac{}{}{0pt}{}{\vec{q}}{j,j`}))_{3r \times 3r}$, where $\alpha, \beta = x, y, z$, and $\vec{q}$ is the wave vector, $j$ and $j`$ denote the atoms in the primitive cell.

a) If $r = 1$, then $j = 1$. The polarization vectors satisfy $\vec{e}_{q,s}(j) \cdot \vec{e}_{q,s`}(j) = \delta_{s,s`}$, $|\vec{e}_{q,s}(j)|^2 = \frac{1}{r} = 1$, $s$

= 1, 2, 3. The dimension of dynamical matrix is (3×3).

b) If $r \geq 2$, the dynamical matrix is a (3$r$×3$r$) Hermitian matrix, which can be partitioned into $r^2$ (3×3) blocks. The same bonding geometry implies, the partial derivatives of potential energy $V$ with respect to atomic displacement $u$, $(\frac{\partial^n V}{\partial u_i \partial u_j \cdots \partial u_k})_0$, are identical for each atom to arbitrary high order $n$, where the subscript 0 denotes the equilibrium position. The matrix elements of $D(\vec{q})$ are obtained by the Fourier transform of the force constants $(\frac{\partial^2 V}{\partial u_i \partial u_j})_0$. As a result, the diagonal blocks of the partitioned matrix $D$ are identical: $(D_{1,1})_{3\times 3} = (D_{2,2})_{3\times 3} = ... = (D_{r,r})_{3\times 3}$. The off-diagonal blocks are complex conjugate of each other: $(D_{i,j})_{3\times 3} = (D_{j,i}^*)_{3\times 3}$. For any two atoms $j$ and $j`$, the corresponding matrix elements in the algebra equations that respectively determine the polarization vectors $\vec{e}_{q,s}(j)$ and $\vec{e}_{q,s}(j`)$ obey the following: $D_{\alpha\beta}(\frac{\vec{q}}{j,j}) = D_{\alpha\beta}(\frac{\vec{q}}{j`,j`})$, $D_{\alpha\beta}(\frac{\vec{q}}{j,j}) = D_{\alpha\beta}^*(\frac{\vec{q}}{j`,j})$. Consequently, $\vec{e}_{q,s}(j)$ and $\vec{e}_{q,s}(j`)$ will differ by only a phase factor, and $|\vec{e}_{q,s}(j)|^2 = |\vec{e}_{q,s}(j`)|^2$. Using the orthonormality $\sum_{j=1}^{r} |\vec{e}_{q,s}(j)|^2 = 1$, one has $|\vec{e}_{q,s}(j)|^2 = |\vec{e}_{q,s}(j`)|^2 = \frac{1}{r}$. Hence the lemma.

*Proof of the theorem.* – Under the geometric condition, it follows from Lemma that the MSD quantities for each atom can be expressed using uniform expressions as given by Eqs. (5)-(7) in the text. As demonstrated by numerous experiments, the VDOS $g(\omega)$ of a crystal takes a parabolic form (Debye model) $g(\omega) = A\omega^2$ at the low frequency region. From Eqs. (5)-(7), it follows that the frequency point $\omega = 0$ is a removable singularity. This is true due to the finite size of realistic systems. Under long wavelength limit, the frequency is $\omega = 2\pi v_s/\lambda$, where $v_s$ is the speed of sound wave. The wavelength satisfies $\lambda \leq L$, with $L$ the dimension of an experimental sample. Hence, $\omega \geq 2\pi v_s/L \equiv \omega_0$, which is minimum frequency. The VDOS $g(\omega) = 0$ for $\omega \leq \omega_0$. The integration interval for Eqs. (5)-(7) can be reduced to $\omega \in [\omega_0, \omega_m]$.

Let $x = \beta\hbar\omega = \hbar\omega/(k_B T)$, the summation $\sum_{n=1}^{\infty}(1-e^{-\beta\hbar\omega})e^{-\beta n\hbar\omega}(n+\frac{1}{2})$ in Eq. (7) turns out to be $\frac{3e^x - 1}{2e^x(e^x - 1)} \equiv f_T(x)$. The expressions of $\langle u^2 \rangle_0$ and $\langle u^2 \rangle_T$ can be rewritten as

$$\langle u^2 \rangle_0 = \int_{\omega_0}^{\omega_m} f_0(x)\frac{\hbar}{m\omega}g(\omega)d\omega, \quad \langle u^2 \rangle_T = \int_{\omega_0}^{\omega_m} f_T(x)\frac{\hbar}{m\omega}g(\omega)d\omega, \text{ with } f_0(x) = (1-e^{-x})/2.$$

$\langle u^2 \rangle_0$ is a continuous, monotonically decreasing function of temperature $T$ while $\langle u^2 \rangle_T$ is the opposite. The lower limit of both $\langle u^2 \rangle_0$ and $\langle u^2 \rangle_T$ is zero. Therefore, there should exit one temperature point at which $\langle u^2 \rangle_0 = \langle u^2 \rangle_T$. Indeed, for each $\omega$, $f_0(x) = f_T(x)$ determines a root that is $x_0 = \hbar\omega/(k_B T_0) = \ln(\frac{5+\sqrt{17}}{2}) \approx 1.518$. Because of the contrary monotonic behavior between $f_0(x)$ and $f_T(x)$ with $x$, $f_0(x) > f_T(x)$ holds for $x > x_0$ while $f_0(x) < f_T(x)$ is true for $x < x_0$. The same relation applies to the integrals that give $\langle u^2 \rangle_0$ and $\langle u^2 \rangle_T$. One can show that $\langle u^2 \rangle_0 > \langle u^2 \rangle_T$ always holds when $T \leq \frac{\hbar\omega_0}{x_0 k_B} \equiv T_1$, while $\langle u^2 \rangle_0 < \langle u^2 \rangle_T$ is always true when $T \geq \frac{\hbar\omega_m}{x_0 k_B} \equiv T_2$. Since $\langle u^2 \rangle_0$ and $\langle u^2 \rangle_T$ are continuous and monotonic, the temperature $T_0$ belongs to the region $T_1 < T_0 < T_2$. This is independent of the form of $g(\omega)$.

**APPENDIX B**

Within the Debye model the VDOS $g(\omega) = A\omega^2$, the two terms are as follows:

$$\langle u^2 \rangle_0 = \frac{A(k_B T)^2}{m\hbar} \int_0^{\theta_D/T} \frac{1}{2}(1-e^{-x})x\,dx \;, \quad \text{and} \quad \langle u^2 \rangle_T = \frac{A(k_B T)^2}{m\hbar} \int_0^{\theta_D/T} \frac{(3e^x-1)}{2e^x(e^x-1)} x\,dx \;.$$

The equality $\langle u^2 \rangle_0 = \langle u^2 \rangle_T$ defines the characteristic temperature $T_0$ as

$$\int_0^{\theta_D/T} \frac{1}{2}(1-e^{-x})x\,dx = \int_0^{\theta_D/T} \frac{(3e^x-1)}{2e^x(e^x-1)} x\,dx. \qquad (B.1)$$

It is clear that $T_0$ is *solely* determined by the Debye temperature $\theta_D$. Our numerical calculation shows that, when the ratio $\zeta = \theta_D/T = 2.99345$, the difference between the left side and right side integral is less than $5\times 10^{-5}$. Therefore, we have

$$T_0 = \frac{\theta_D}{\zeta} \approx \frac{\theta_D}{2.99345}. \qquad (B.2)$$

Table I. Calculated total MSD $\langle u^2 \rangle$ of Al and diamond at different temperatures, by using DFPT and Debye VDOS.

|  | Al | | | Diamond | | |
|---|---|---|---|---|---|---|
| $T$ (K) | 4.2 | 77 | 293 | 4.2 | 77 | 293 |
| [a]DFPT (Å$^2$) | 0.0096 | 0.0113 | 0.0268 | 0.0048 | 0.0048 | 0.0053 |
| [b]Debye (Å$^2$) | 0.0095 | 0.0114 | 0.0274 | 0.0041 | 0.0041 | 0.0045 |
| [c]Debye (Å$^2$) | 0.0095 | 0.0114 | 0.0274 | 0.0042 | 0.0042 | 0.0046 |

[a]Present work.

[b]Present work.

[c]The MSD is deduced by using the relation DWF = $\frac{8\pi^2}{3} \langle u^2 \rangle$ for an isotropic crystal, where the DWF data are from Reference [10].

Table II. The Debye temperature ($\theta_D$), characteristic temperature ($T_0$) of zero-point vibration, and superconducting transition temperature ($T_c$) of a number of elements. The quantity $\eta(T) = \langle u^2 \rangle_0 / \langle u^2 \rangle$. The data of $\theta_D$ (low temperature limit) and $T_c$ are from Ref. [18]. The unit for temperatures is K.

| Element | $\theta_D$ | $T_0$ | $T_c$ | $\eta(T_c)$ | $\theta_D/T_0$ |
|---|---|---|---|---|---|
| Be | 1440 | 481.06 | 0.026 | 1.000 | 2.993 |
| Al | 428 | 142.98 | 1.140 | 1.000 | 2.993 |
| Ti | 420 | 140.31 | 0.39 | 1.000 | 2.993 |
| Ta | 240 | 80.18 | 4.483 | 0.997 | 2.993 |
| Pb | 105 | 35.08 | 7.193 | 0.961 | 2.993 |

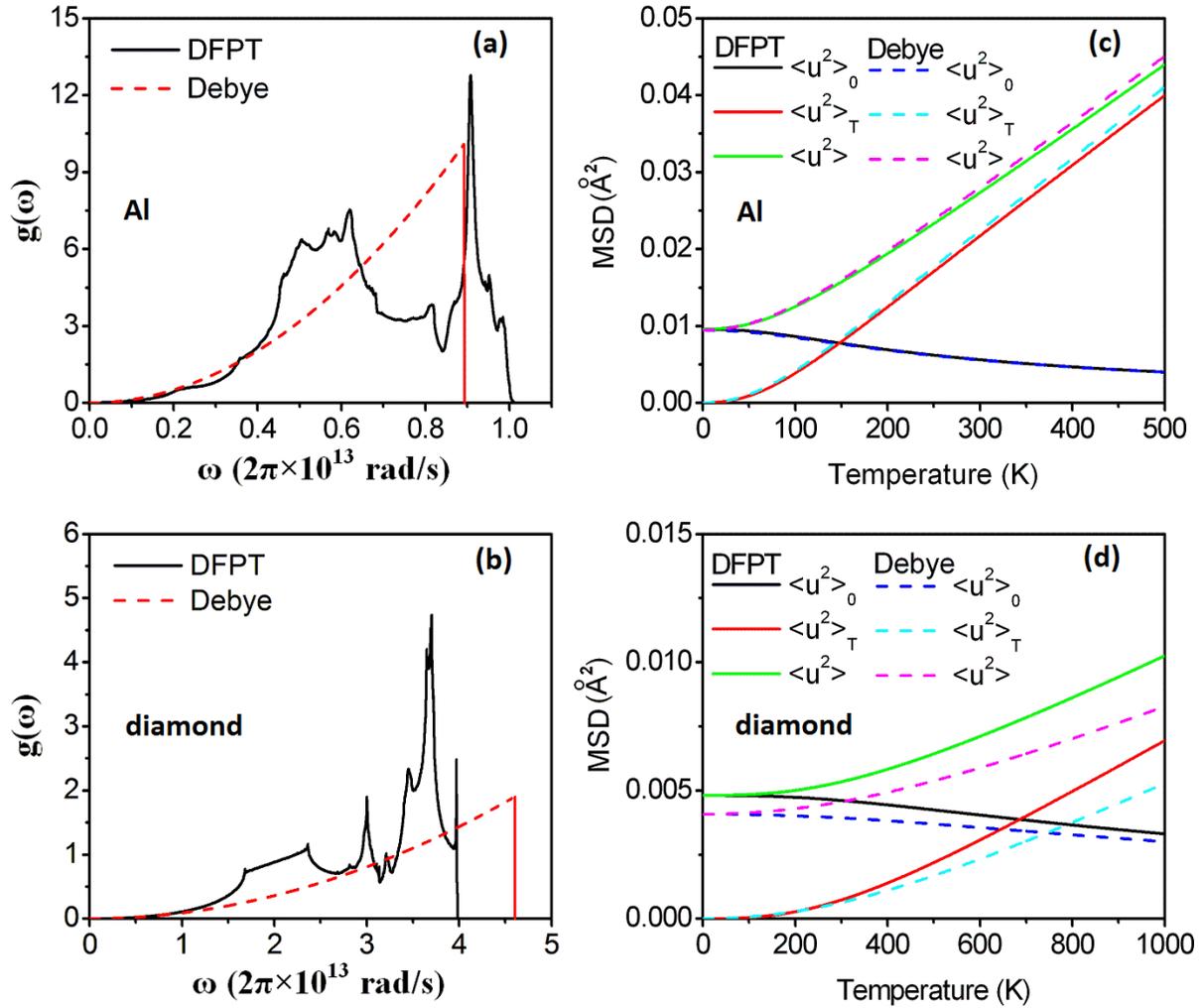

FIG. 1 (color online) **Panels (a)-(b):** Vibrational density of states ($g(\omega)$) of Al and diamond, calculated from the density functional perturbation theory (DFPT) and from the Debye model (dash lines). **Panels (c)-(d):** Calculated atomic mean square displacement (MSD) in Al and diamond, as a function of temperature, by using the $g(\omega)$ of DFPT (solid lines) and the Debye model (dash lines).